\documentclass[preprint,showpacs,showkeys,preprintnumbers,amsmath,amssymb]{revtex4}

\usepackage{graphicx}% Include figure files
\usepackage{dcolumn}% Align table columns on decimal point
\usepackage{bm}% bold math

\begin{document}

\preprint{}

\title{Extended Fractal Fits to Riemann Zeros}

\author{Paul B. Slater}% 
\email{slater@kitp.ucsb.edu}
\affiliation{%
ISBER, University of California, Santa Barbara, CA 93106\\
}%
\date{\today}% It is always \today, today,

\begin{abstract} 
We extend to the first 300 Riemann zeros, the form of analysis
reported by us in Canad. J. Phys. {{\bf 85}}, 1 (2007) 
(also arXiv:math-ph/0606005), in which the largest study 
had involved 
the first 75 zeros.
Again, we model 
the nonsmooth fluctuating part of the Wu-Sprung potential, which reproduces
the Riemann zeros, by the alternating-sign sine series 
fractal of Berry and Lewis $A(x,\gamma)$. Setting the fractal dimension 
equal to $\frac{3}{2}$, 
we estimate the 
frequency parameter ($\gamma$), plus an overall scaling 
parameter ($\sigma$) introduced.
We search for that pair of parameters ($\gamma,\sigma$) 
which {\it minimizes} the least-squares
fit of 
the lowest 300 eigenvalues --- obtained by solving the
one-dimensional stationary (non-fractal)
Schr\"odinger equation with the trial potential (smooth {\it plus} 
nonsmooth parts) --- to the 
first 300 Riemann zeros. 
We randomly sample values within the rectangle
$0 < \sigma < 3, 0<\gamma <25$.
The fits obtained are 
compared to those gotten by  using simply the {\it smooth}
part of the Wu-Sprung potential {\it without} any fractal
supplementation. Some limited improvement is again
found. There are two (primary and secondary) 
quite distinct subdomains, in which the values giving improvements 
in fit are concentrated.
\newline
\newline
\end{abstract}


\maketitle

\section{Introduction}
Our objective here is the discernment of  
structural patterns within the nontrivial
Riemann zeros (that is, the real parts $\alpha_{k}>0$ of the zeros 
of the Riemann zeta function $\zeta(s)$, which 
conjecturally all take the form $\frac{1}{2}+ \alpha_{k} i$).
More specifically, following our work in \cite{slaterCanadian} 
(also arXiv:math-ph/0606005), 
we seek to model the fractal component of the
potential which reproduces the $\alpha_{k}$'s.
The (dominant) smooth part of such a potential can be written 
implicitly in the form,
\begin{equation} \label{WSpotential}
x_{WS}(V)=\frac{1}{\pi} 
\Big( \sqrt{V-V_{0}} \ln \frac{V_{0}}{2 \pi e^2} +  \sqrt{V} 
\ln \frac{\sqrt{V}+\sqrt{V-V_{0}}}{\sqrt{V}-\sqrt{V-V_{0}}} \Big).
\end{equation}
(Here $V_{0}= 3.10073 \pi \approx 9.74123$.)
This formula (\ref{WSpotential}) 
was derived by Wu and Sprung \cite{wu} by solving a form
of Abel's integral equation obtained by the application of the 
semiclassical approximation to the  formula,
\begin{equation}
N(E)=\frac{E}{2 \pi} \ln{\frac{E}{2 \pi E}} +\frac{7}{8},
\end{equation}
for the smooth part of the number of 
Riemann zeros less than $E$ (cf. \cite{garcia}).

In \cite{slaterCanadian}, we modeled the fractal
component by the alternating-sign sine series fractal 
(a particular case of a deterministic Weierstrass-Mandelbrot fractal 
function) of Berry
and Lewis \cite[eq. (5)]{berry},
\begin{equation} \label{alternating}
A(x,\gamma) =\Sigma_{m=-\infty}^{\infty}
\frac{(-1)^m \sin{\gamma^m x}}{\gamma^{(2-d) m}}, \hspace{1in}                 \
(1 <d  < 2, 1 <  \gamma).
\end{equation}
Here, $d$ is the fractal dimension, which --- following
the  box-counting argument of
Wu and Sprung \cite{wu}
(cf. \cite{vanZyl}) --- we took to be $\frac{3}{2}$. (van Zyl and 
Hutchinson also came to the same conclusion regarding the fractal 
dimension, independently of
which of 
two distinct inversion techniques they employed \cite{vanZyl}.)
We have, in this $d=\frac{3}{2}$ Berry-Lewis context,  a specific case,
\begin{equation} \label{AffineScaling}
A(\gamma x,\gamma)= - \gamma^{\frac{1}{2}}  A(x,\gamma),
\end{equation}
of the  ``affine scaling law'' \cite[eq. (3)]{berry}.

In the most extensive of the three ($n =25, 50, 75$) analyses  
reported in \cite{slaterCanadian},
we found that values of the frequency parameter 
$\gamma$ in the vicinity of 3 (and, secondarily, 9)
tended to provide the best fits to the Riemann zeros 
($\alpha_{k}$'s) themselves \cite[Fig.~16]{slaterCanadian}.
(The sum of the inverted Wu-Sprung potential (\ref{WSpotential}) plus
the alternating-sign sine series (\ref{alternating}), 
with $\gamma$ fixed at various 
random values,  was used in these analyses as an adjusted potential, 
which served as input to the time-independent 
one-dimensional Schr{\"o}dinger equation. The eigenvalues obtained  are, then, 
regarded as fits to the corresponding Riemann zeros.)

Here, we continue along these same investigative 
lines, quadrupoling the sample size 
of zeros 
from 75 to 300. We achieve this 
substantial increase mainly through reducing --- from 
5,625 to 600 --- the basis size employed in the Arnoldi method for
finding eigenvalues of sparse matrices 
 as well 
as sampling 6,500 points, not 10,000, in the Mathematica program 
supplied to us by M. Trott. 
(We also truncate the series (\ref{alternating}) at $m=40$, while in
\cite{slaterCanadian} we used a cutoff of $m=30$ --- which, of course, 
does not reduce computational time.)
Despite these reductions, our program, when we  employ just
the smooth (unsupplemented) Wu-Sprung potential (\ref{WSpotential}),
yielded what we deemed to be a suitably acceptable
prediction of 542.023 for the (highest)
300-{\it th} Riemann zero, that is, 541.847.
( It would certainly 
be of interest to evaluate the sensitivity of our results to the various
computer program parameters employed and adjust them
accordingly, but it would seem to be 
overwhelmingly time-consuming to do so in substantial detail. 
We also tried to adopt our Mathematica program to analyze the first 500 
Riemann zeros, but we could not adjust the parameters to provide us with
an acceptable fit, as measured by comparing the 500-{\it th} 
eigenvalue with the 500-{\it th} Riemann zero.)

While, in the $n=75$ analysis in 
\cite{slaterCanadian}, we randomly sampled values of $\gamma 
\in [0,10]$, here we expand the interval to $\gamma \in 
[0,25]$, and as in the first of
the analyses ($n=25$) in \cite{slaterCanadian}, allow the series 
(\ref{alternating})
to be scaled by an overall factor of $\sigma$, we take now to lie  in 
the interval [0,3].
(We arrived at this rectangular range of parameters from which to 
extensively sample, after a number of preliminary analyses.)

\begin{figure}
\includegraphics{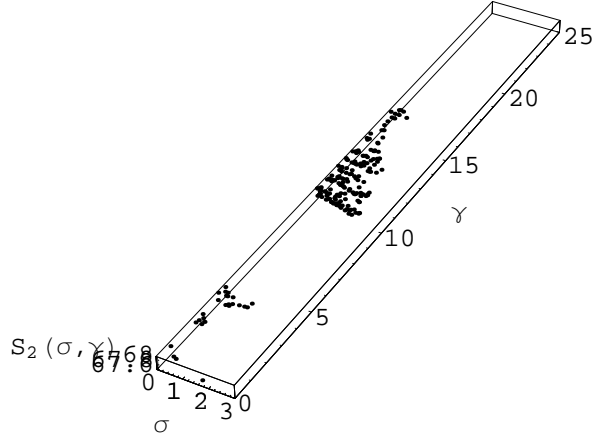}
\caption{\label{fig:Boxy}Those 100 pairs ($\sigma,\gamma$) which give
improved alternating-sign sine series (\ref{alternating}) 
fits (for all $k=1,\frac{3}{2}, 2$) 
to the first 300 Riemann zeros. 20,000 pairs were randomly,
{\it uniformly} sampled over the base rectangle 
($0 < \sigma <3, 0< \gamma <25$). The third axis $S_{2}(\sigma,\gamma)$ 
gives the 
associated least-square fit ($j=2$) --- smaller values, of course, denoting
better fits to the zeros. The minimum value of 67.562 is attained at
(10.4444, 1.50996).}
\end{figure}

In Fig.~\ref{fig:Boxy} we show the results obtained by randomly sampling
points in the indicated base two-dimensional
rectangle and determining how well the so-supplemented
smooth Wu-Sprung potential fits the first 300 Riemann zeros. Only those
213 points (of the 3,330 sampled) 
for which the fit is improved for all three measures
$S_{k}(\sigma,\gamma)  = 
\Sigma_{k=1}^{300} |\tilde{\alpha_{k}}-\alpha_{k}|^j, j =1, \frac{3}{2}, 2$ 
are displayed. (The $\tilde{\alpha}_{k}$'s denote, say, the eigenvalues, 
and the $\alpha_{k}$'s the Riemann zeros.) 
These measures were, respectively, 116.265, 86.3596 and 
68.1307 when no fractal component was added, and only the smooth potential
employed. So, Fig.~\ref{fig:Boxy} contains only 
points for which all three indicators are less than these three values.
All the improvements are rather minor in character --- as 
in \cite{slaterCanadian} --- with none exceeding
one-percent in terms of the criterion values $S_{k}(\sigma,\gamma)$.
We observe that for most of these points yielding improvements in fit, 
we have  $\gamma>10$, a region that was 
outside the range of consideration in \cite{slaterCanadian}.
The minimum values using the $j=1$ and $j= \frac{3}{2}$ 
criteria, 115.451 and 85.6691, were attained at the same point
$\gamma=10.3954,\sigma=1.55535$, 
while using the $j=2$ (least-squares) criterion, the minimizing point, 
with the value 67.562, was 
located at $\gamma=10.4444,\sigma=1.50996$.
(For all of the 213 improving values, we had $\sigma< 1.85799$.)

\begin{figure}
\includegraphics{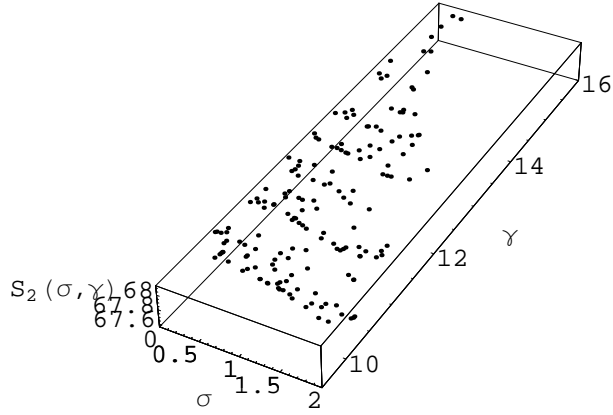}
\caption{\label{fig:RiemannBox2}Magnification of 
primary  domain of fit-improving points in Fig.~\ref{fig:Boxy}}
\end{figure}

\begin{figure}
\includegraphics{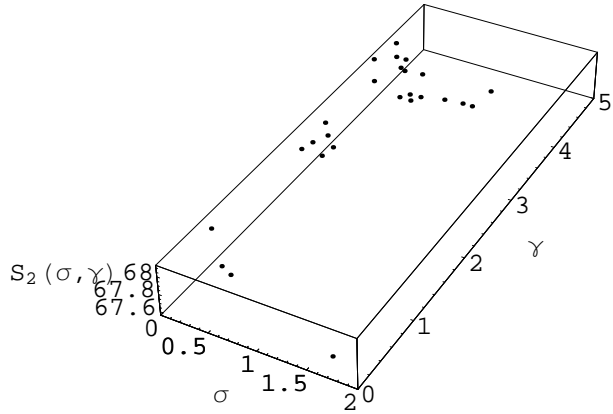}
\caption{\label{fig:RiemannBox3}Magnification of 
secondary domain of fit-improving points in Fig.~\ref{fig:Boxy}}
\end{figure}

In Figs.~\ref{fig:RiemannBox2} and \ref{fig:RiemannBox3}, 
we magnify the regions in 
Fig.~\ref{fig:Boxy}, in which the improvements in fit are
concentrated.

We also investigated the possibility of modifying  our analyses by 
adding a prefactor of $\gamma^{-\frac{x}{2}}$ to (\ref{alternating}) 
to enhance periodicity, 
following certain suggestions of Tricot \cite[sec. 12.11]{tricot}, 
but this led to convergence  problems, which we did not know immediately how
to overcome. Of course, it would be of interest to employ still other 
basic (few-parameter)
fractal models in addition to 
(\ref{alternating}) in similar analyses of the
fractal component of the Wu-Sprung potential for the Riemann zeros, 
possibly thereby achieving greater improvements in fits to the 
zeros than we have so far been able to obtain.
(In this regard, the companion 
cosine series of Berry and Lewis \cite[eq. (4)]{berry} 
is of potential interest, although unlike (\ref{alternating}), it is 
never negative, so it would seem necessary to add some negative quantity 
to it.)

It would also be of interest to directly model the fractal
component 
remaining when the smooth Wu-Sprung potential is fitted to the Riemann zeros
(cf. \cite[inset, Fig. 2]{wu} \cite[Fig. 3 (a)-(c)]{vanZyl}), 
rather than having --- as we have done here and in 
\cite{slaterCanadian} ---  inputting each candidate for 
the fractal component into the time-independent one-dimensional 
Schr{\"o}dinger equation.

\begin{acknowledgments}
I would like to express gratitude to the Kavli Institute for Theoretical
Physics (KITP)
for computational support in this research.
\end{acknowledgments}

\bibliography{NewRiemann}% Produces the bibliography via BibTeX.

\end{document}